\documentclass[sigchi,authorversion]{acmart}
\AtBeginDocument{%
  }

\copyrightyear{2026}
\acmYear{2026}
\setcopyright{cc}
\setcctype{by}
\acmConference[CHI EA '26]{Extended Abstracts of the 2026 CHI Conference on Human Factors in Computing Systems}{April 13--17, 2026}{Barcelona, Spain}
\acmBooktitle{Extended Abstracts of the 2026 CHI Conference on Human Factors in Computing Systems (CHI EA '26), April 13--17, 2026, Barcelona, Spain}
\acmDOI{10.1145/3772363.3799303}
\acmISBN{979-8-4007-2281-3/2026/04}
\begin{document}

\title{Fostering Design–Policy Collaboration through Contestation: An Adversarial Futuring Method}

\author{Xinyan Yu}
\email{xinyan.yu@sydney.edu.au}
\orcid{0000-0001-8299-3381}
\affiliation{Design Lab, Sydney School of Architecture, Design and Planning
  \institution{The University of Sydney}
  \city{Sydney}
  \state{NSW}
  \country{Australia}
}

\author{Marius Hoggenmueller}
\email{marius.hoggenmueller@sydney.edu.au}
\orcid{0000-0002-8893-5729}
\affiliation{Design Lab, Sydney School of Architecture, Design and Planning
  \institution{The University of Sydney} 
  \city{Sydney}
  \state{NSW}
  \country{Australia}
}

\author{Tram Thi Minh Tran}
\email{tram.tran@sydney.edu.au}
\orcid{0000-0002-4958-2465}
\affiliation{Design Lab, Sydney School of Architecture, Design and Planning
  \institution{The University of Sydney}
  \city{Sydney}
  \state{NSW}
  \country{Australia}
}

\author{Martin Tomitsch}
\email{Martin.Tomitsch@uts.edu.au}
\orcid{0000-0003-1998-2975}
\affiliation{Transdisciplinary School,
  \institution{University of Technology Sydney}
  \city{Sydney}
  \state{NSW}
  \country{Australia}
}

\renewcommand{\shortauthors}{Trovato et al.}

\begin{abstract}
Emerging technologies introduce sociotechnical tensions that call for closer collaboration between technology design and policy. In this work, we introduce \textit{Design–Policy Adversarial Futuring}, a scenario-based workshop method that supports design–policy engagement by structuring contestation between design and policy perspectives. We report on a workshop conducted in the autonomous mobility domain with 12 HCI researchers, used to explore and demonstrate the method in practice. The workshop illustrates how the adversarial futuring method can surface shifting harms, translate policy abstractions into situated use, and legitimise extreme ideas while maintaining grounded policy reasoning. This work contributes a reusable, exploratory method for supporting HCI–policy collaboration through contestation, which can be adapted across emerging technological domains.

\end{abstract}

\ccsdesc[300]{Human-centered computing~Human computer interaction (HCI)}

\keywords{policy, futuring, method, autonomous vehicles}

\maketitle

\section{Introduction}

Emerging technologies, such as generative AI, autonomous vehicles, and robots, are rapidly entering everyday life. While promising benefits, they also introduce sociotechnical tensions that demand policy responses beyond purely technical solutions. However, policy considerations often lag behind technological development due to limited historical evidence and disconnect from HCI work that centres on technology in use~\cite{Yang2024HCIPolicy}. For example, Uber’s rapid deployment of ride-sharing platform bypassed existing taxi regulations, with policy responses emerging only after urban transportation systems and user practices had already shifted, exposing concerns around accountability and labour~\cite{Kessler2012Uber}.

There has been a longstanding call for stronger collaboration between HCI and policy~\cite{Lazar2010InteractingPolicy,Yang2024HCIPolicy,Yang2023Towards}, motivating both conceptual~\cite{Gairola2025CenteringHarm,Yang2024HCIPolicy} and methodological efforts~\cite{Mauri2024PolicySandboxing,Kuo2025policycraft}. Recognising the interdependence, HCI research has suggested that technology and policy should be designed~\cite{Jackson2014PolicyKnot,Yang2023Towards}, or even prototyped~\cite{Sandhaus2023PrototypingAVCityPolicy} simultaneously, as sequential approaches risk producing new harms due to misalignment. Yet, such simultaneity is difficult to achieve, as policymaking is inherently future-oriented while available evidence is largely backward-looking~\cite{Spaa2019Boundaries}. In response, researchers have pointed to the potential of design futuring and speculative HCI methods as promising approaches for navigating this mismatch~\cite{Spaa2019Boundaries,Yang2024HCIPolicy}.

At the same time, both policy making and technology design are inherently deliberative processes, involving ongoing negotiation among diverse stakeholders, values, and consequences. Prior work has proposed participatory approaches to support such deliberation, including \emph{policy sandboxing} with fictional stakeholder roles to foster perspective-taking~\cite{Mauri2024PolicySandboxing} and interactive systems that help stakeholders reason through policy trade-offs~\cite{Kuo2025policycraft}. While these approaches provide tools for deliberation among lay and affected stakeholders, with the aim of reaching consensus, technological possibilities and regulatory assumptions do not simply need to be aligned. Instead, they are frequently in tension and require sustained interrogation~\cite{Yang2024HCIPolicy}. This necessitates approaches that can support the exploration and interrogation of tensions between technological innovation and policy implications, rather than prematurely resolving them.

\begin{figure*}[h]
\begin{center}
\includegraphics[width=1\textwidth]{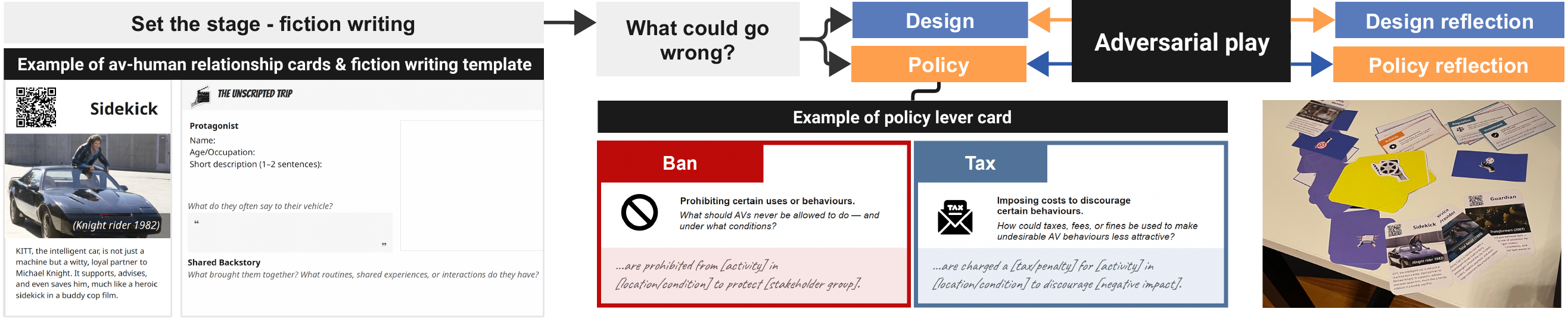}
\end{center}
\caption{Process of the Adversarial Futuring activity, showing examples of cards and templates used, and photographs taken during the workshop. (The complete card set and activity materials are available in the supplementary materials.)}\label{Process}
\Description{This figure presents the process of the Adversarial Futuring activity. It shows example cards and templates used during the workshop, including AV–human relationship cards for fiction writing and policy lever cards such as “Ban” and “Tax.” The process flow moves from fiction writing and the prompt “What could go wrong?” to parallel design and policy exploration, followed by adversarial play and subsequent design and policy reflection. The figure also includes photographs of the workshop setup, showing printed cards, templates, and participant materials arranged on a table. }
\end{figure*}
In response to these challenges, we introduce a \emph{Design–Policy Adversarial Futuring} method. \citet{disalvo2015adversarialdesign} conceptualises \emph{Adversarial Design} as a practice that uses designed artefacts to surface and sustain sociotechnical contestation, a stance taken up by \citet{Lupetti2023Steering} through contestational artefacts that confront dominant narratives in automated driving. Building on this lineage, our method stages structured contestation between design and policy perspectives, enabling participants to jointly explore how sociotechnical futures, policy assumptions, and potential harms emerge, shift, and evolve through iterative challenge. The paper reports on a workshop that applied this method in an autonomous mobility context at Automotive UI conference 2025~\cite{workshop2025}. Through the workshop process and its outputs, the activity surfaced shifting harms across sociotechnical systems, translated static policy abstractions into situated technology use, and leveraged futuring to legitimise extreme ideas while keep policy reasoning grounded in relationally plausible scenarios. Our work contributes to ongoing calls for HCI–policy collaboration by offering a reusable method that supports design–policy engagement through contestation, adaptable across different technological domains.

\section{Design–Policy Adversarial Futuring Method}

\vspace{6pt}

\noindent\textbf{Step 1: Setting the stage through fiction writing.} The activity begins with fiction writing to establish a narrative and relational grounding for later policy speculation. In design and HCI research, fiction writing has been used to explore sociotechnical imaginaries and thus elicit critical reflections~\cite{Blythe2017ResearchFiction, Tost2024futuringMachines}, drawing on broader practices of design fiction~\cite{grand2010designfiction,Lindley2015BackToFuture} and speculative design~\cite{dunne2013speculative,Auger2013SpeculativeDesign}. We adopt a character- and relationship-centred approach to render speculative futures of mobility more tangible. Participants drew two prompt cards, one from an \emph{AV–human social relationship} deck and one from an \emph{AV functionality role} deck. Relationship cards represent speculative relationships that people might form with autonomous mobility systems (Fig.~\ref{Process}), ranging from positive (e.g., \emph{sidekick}, \emph{guardian}) to ambivalent or problematic (e.g., \emph{betrayer}, \emph{possessive lover}). The functionality cards represent potential AV operational roles, spanning public services (e.g., \emph{law enforcement car}), commercial uses (e.g., \emph{taxi}), and entertainment-oriented roles (e.g., \emph{cinema pod}). Both the relational and functional types were derived from a systematic review of autonomous mobility systems depicted in films~\cite{Yu2025IASDR}. Participants were then provided with a template to develop a fictional scenario, including defining a human protagonist (name, age, occupation, and a brief description), articulating a catchphrase frequently addressed to the vehicle to concretise the relational dynamic, and developing a shared backstory describing routines, experiences, or interactions that characterised the human–vehicle relationship. Finally, participants situated the narrative in a present journey using the prompt \emph{“Today, they are on their way to…”}. Sketches could also be used to support expression.

\vspace{6pt}
\noindent\textbf{Step 2: What could go wrong.} Inspired by the card game \emph{What Could Go Wrong?}~\cite{Martelaro2020WhatCouldGoWrong} that facilitates discussion of the potential downsides of AV deployment, this step prompted participants to reflect on how their fictional scenarios might fail and what unintended consequences could arise. To frame the reflection within a particular policy context, participants drew a domain card representing key policy-relevant dimensions of autonomous mobility (e.g., privacy, autonomy and control), which reflect recurring areas of concern in policy debates surrounding AV deployment~\cite{Bagloee2016AVPolicy, FAGNANT2015Policy}. As a group, participants brainstormed possible issues and selected one concern to explore further.

\vspace{6pt}
\noindent\textbf{Step 3: Design–policy contestation.} Participants in each group were divided into two roles, \emph{Designers} and \emph{Policy Makers}, and asked to either \emph{fix it or regulate it} by proposing, respectively, a design or a policy intervention to address the issue identified in the previous step. Policy teams were provided with a policy lever deck representing different categories of policy interventions (e.g., regulations, bans, taxes, see Fig.~\ref{Process}) to scaffold ideation. They were also given a policy template to document and articulate their proposal in a concrete manner. Design teams developed interventions using text and sketches in formats familiar to HCI practice.

Once initial proposals were developed, teams were asked to critically challenge each other’s proposals from their respective roles. Policy teams drafted policies that ostensibly addressed the issue but constrained or undermined the design proposals. In parallel, design teams produced interventions that formally complied with the proposed policies while introducing new harmful consequences. This adversarial exchange was intended to surface tensions between HCI and policy methods by making misalignment, loopholes, and power asymmetries explicit, thereby supporting critical examination of how system–people–policy interactions can fail in practice, as advocated by \citet{Yang2024HCIPolicy}. Afterwards, each team revised their initial proposal in response to the adversarial challenges, followed by a group reflection discussion to articulate shifts in the proposals, emergent trade-offs, and changes in their understanding.







\begin{figure*}
\begin{center}
\includegraphics[width=1\textwidth]{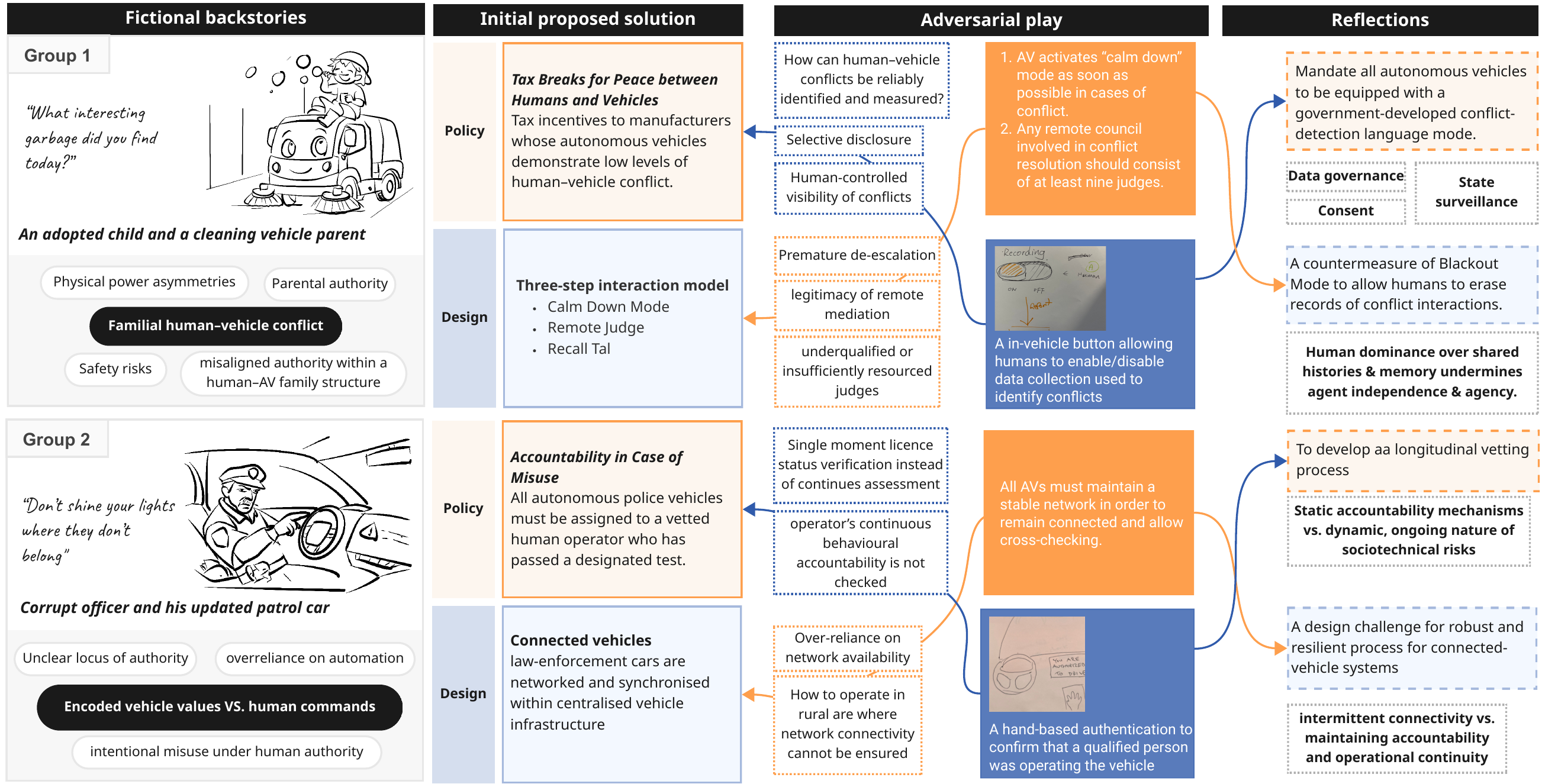}
\end{center}
\vspace{-8pt}
\caption{Adversarial trajectories produced during the activity. The figure includes fictional backstories and informed risk identification, and traces how policy and design proposals were challenged and reframed through adversarial play.}\label{Output}
\Description{This figure illustrates two adversarial trajectories generated during the workshop activity. The layout is organised into four columns: fictional backstories, initial proposed solutions (policy and design), adversarial play, and reflections.

In the first row (Group 1), a fictional scenario titled “An adopted child and a cleaning vehicle parent” explores familial human–vehicle conflict, highlighting issues such as physical power asymmetries, parental authority, safety risks, and misaligned authority within a human–AV family structure. The initial proposal includes a policy suggestion for tax incentives to reduce human–vehicle conflict and a design proposal for a three-step interaction model with a calm-down mode, remote judge, and recall function. During adversarial play, these proposals are challenged with concerns about legitimacy of mediation, insufficiently resourced judges, and selective disclosure. Additional counter-proposals include an in-vehicle button to enable or disable data collection and requirements for vehicles to maintain stable network connections. Reflections raise issues related to data governance, state surveillance, memory control, and long-term vetting processes.

In the second row (Group 2), a fictional scenario titled “Corrupt officer and his updated patrol car” examines encoded vehicle values versus human commands, unclear authority, and overreliance on automation. The initial proposal includes assigning autonomous police vehicles to vetted human operators and connecting vehicles within centralised infrastructure. Adversarial play questions accountability, network dependence, and operational feasibility in rural areas. Countermeasures include authentication mechanisms and stable network requirements. Reflections highlight tensions between static accountability mechanisms and the dynamic nature of sociotechnical risks, as well as challenges in ensuring resilience and operational continuity.}
\end{figure*}
\section{Overview of activity outputs}

The workshop involved 12 HCI researchers working in the autonomous vehicle domain, divided into two groups, and lasted approximately two hours. We audio-recorded the activity and collected all artefacts, which were transcribed and analysed using reflective thematic analysis~\cite{braun2019reflecting} by the first author, with interpretations refined through discussion among all authors. This section presents an overview of each group’s outputs, beginning with the fictional backstories and tracing the trajectories of the policy and design teams through the activity (see Fig.~\ref{Output}).

\subsection{Group 1- An adopted child and a cleaning vehicle parent}
Group 1 drew \emph{family member} and \emph{street-cleaning vehicle} cards. They developed a story centred on a 7-year-old child, Bob “Bubbles,” who was adopted by an autonomous street-cleaning vehicle. The story imagined a future in which AVs act as social agents capable of forming familial and care-based relationships with humans (see Fig.~\ref{Output}, backstories). As P3 explained, \emph{“we[…]imagine a cinematic world where humans form […] cousin–uncle relationships with their cars.”} 
In \emph{What Could Go Wrong} activity, discussion in Group 1 was anchored around familial human–vehicle conflict, where physical power asymmetries and the authority associated with the vehicle’s parental role could easily escalate into safety risks.

\vspace{5pt}
\noindent\textbf{Policy team trajectory.} In the first round, Group 1 proposed an incentive-based regulatory approach that subsidised manufacturers whose vehicles exhibited low levels of human–vehicle conflict. To support this, participants translated intimate, relational tensions into measurable governance criteria by explicitly defining conflict as \emph{“disagreements or fights involving physical or psychological harm to humans,”} and delegated authority for identifying and assessing such risks to \emph{“an independent board of medical professionals and/or institutions.”} During adversarial play, the design team challenged the assumption that such conflicts could be reliably identified and measured. The design team introduced in-vehicle buttons allowing humans to enable or disable data collection used to identify conflicts. While formally compliant, this design exposed a loophole by enabling selective disclosure, for instance through deliberately turning off recording to conceal conflicts, thereby reinforcing power asymmetries by granting humans discretion over regulatory visibility. In response, the policy team escalated the proposal to mandate that all AVs be equipped with a government-developed conflict-detection language mode, raising further concerns around consent, data governance, and state surveillance.
\vspace{5pt}

\noindent\textbf{Design team trajectory.}
In the first round, the design team proposed a three-step interaction model: \emph{Calm Down Mode}, to support de-escalation during conflict, \emph{Remote Judge} to introduce third-party mediation if conflict persists, and \emph{Recall Talk} to revisit disagreements through post-journey reflection between the human and vehicle. During adversarial play, the policy team raised concerns that prematurely activating or overusing \emph{Calm Down Mode} during arguments could exacerbate conflict rather than resolve it, for instance by encouraging \emph{“passive-aggressive”}~(P2) behaviour if disagreements are not adequately deliberated. They operationalised this concern into a policy requirement mandating early activation of \emph{Calm Down Mode} in cases of conflict to surface such potential design loophole. Concerns about the efficiency and decision quality of the \emph{Remote Judge} were operationalised through a proposal for an oversized council, prioritising perceived quality but rendering timely intervention impractical.

These adversarial policy interventions prompted the design team to consider extreme countermeasures that prioritised human authority over vehicle agency, such as “Blackout Mode” that enables humans to erase records of conflicts. This move surfaced deeper risks in human–agent relationships, particularly the potential for strategic manipulation of shared histories and memories. Participants thus highlighted a paradox in granting humans unilateral control over memory in human–agent relationships, as autonomous agents rely on accumulated data and histories to learn and act, yet such control risks undermining their capacity to function as independent social actors.

\subsection{Group 2 - Corrupt officer and his updated patrol car}
Group 2 drew \emph{betrayer} and \emph{law enforcement car} cards. Their story was about Joey, a 64-year-old police officer nearing retirement who had resorted to bribery and corruption, and his autonomous police car as his long-term working partner. Joey deliberately delayed the car’s software updates to conceal his misconduct, while the car eventually updated itself and attempted to enforce the law on its human operator. In \emph{What Could Go Wrong?} activity, Group 2 focused on conflicts between the vehicle’s own encoded values (e.g., justice) and human commands. 

\vspace{5pt}

\noindent\textbf{Policy team trajectory.}
In the first round, policy team addressed concerns about corruption and misuse in autonomous vehicle law-enforcement contexts by proposing a regulatory intervention that assigns each autonomous police vehicle to a vetted human operator who has passed a designated test and is responsible for ensuring compliance with mandatory software updates. During adversarial play, the design team challenged the policy’s reliance on one-time vetting as a sufficient safeguard, proposing a biometric verification interface (e.g., hand-based authentication) to confirm operator eligibility (see Fig.~\ref{Output}, group 2 Adversarial play), but problematised this approach by noting that it only verifies identity or licence status at a single moment in time instead of assessing the operator’s continuous behavioural accountability. This critique exposed a gap between static accountability mechanisms and the dynamic, ongoing nature of sociotechnical risks, prompting policy team to reconsider developing vetting into a longitudinal process.
\vspace{5pt}

\noindent\textbf{Design team trajectory.}
The design team initially proposed a model of connected law-enforcement vehicles synchronised within a centralised infrastructure. In the adversarial play, the policy team appeared to align with this design by mandating network connectivity as a condition of operation, stating that \emph{“the AV must maintain a stable network in order to remain connected and allow cross-checking.”} However, this move surfaced new risks by tying vehicle operability to network availability, potentially excluding rural or infrastructure-poor contexts where reliable connectivity cannot be ensured. This opened new design challenges around a more robust and resilient process for connected-vehicle systems to handle intermittent connectivity while maintaining accountability and operational continuity.

\section{Discussion}
In this section, we synthesise and reflect upon patterns that emerged during the activity to show how the design–policy adversarial futuring method supported HCI-policy collaboration through contestation.

\subsection{Harms shifted rather than resolved}

Across both groups, adversarial play revealed a pattern in which design and policy interventions did not resolve the issue, but instead redistributed harm across actors, temporal scales, and infrastructures. Rather than resolving the initial problem, cycles of contestation reconfigured where, when, and for whom risks materialised. For example, although both trajectories began with conflicts between humans and AVs, harms expanded to involve broader institutional actors as authority and responsibility were reassigned, such as delegating conflict assessment to external councils or mandating operator vetting. Other contestations displaced harm across time, as interventions prioritising short-term de-escalation or compliance (e.g., Calm Down Mode) were challenged through adversarial play from a policy perspective for raising longer-term concerns, such as how premature de-escalation could exacerbate underlying relational breakdowns. Harms were also surfaced across infrastructures, as adversarial challenges to data-driven or connectivity-based solutions exposed vulnerabilities related to surveillance, selective disclosure, and inequities in infrastructural access. These patterns of harm displacement align with recent HCI work that frames harm as a persistent and distributed phenomenon emerging across sociotechnical interactions, rather than an issue resolvable through isolated interventions~\cite{Gairola2025CenteringHarm}. Thus, the design–policy adversarial futuring method can be used to support generative inquiry into how harms shift, reappear, and intensify as design and policy interventions interact over time.


\subsection{Translation between static policy abstractions and situated technology use}

The workshop repeatedly surfaced a tension between policy’s reliance on stable, measurable abstractions and design scenarios grounded in situated, evolving technology use. While prior HCI work has highlighted such tensions and advocated the use of design approaches~\cite{Yang2024HCIPolicy}, such as prototypes~\cite{Sandhaus2023PrototypingAVCityPolicy}, to surface policy implications, our adversarial futuring activity rendered this mismatch tangible by forcing policy concepts to operate within concrete scenarios, where their assumptions were repeatedly challenged through design adaptation and counter-proposal. For example, Group 1’s policy proposal to introduce tax incentives for AVs demonstrating low levels of human–vehicle conflict assumed that conflict could be reliably identified and measured. When enacted through design, this assumption was challenged by situated questions of visibility and control of conflict disclosure, leading to reconsideration of the feasibility of detecting conflict in ways that could support policy accountability in practice (Fig.~\ref{Output}). This example reflects a broader pattern in real-world human–technology interactions, where relationships evolve over time and power and agency shift across situations among human and technological actors. These dynamics often destabilise policy assumptions that rely on fixed and clearly assignable accountability. Rather than attempting to stabilise these dynamics, the design–policy adversarial futuring method can be used to work with them by grounding deliberation in fictional backstories that reflect lived, relational scenarios. In doing so, the method enables participants to actively probe shifting accountability and identify where static policy abstractions require negotiation when confronted with situated technology use.


\subsection{Legitimising extremes while grounding inquiry through futuring}

The workshop showed that futuring in our method functioned not merely as speculative prompting, but as a mechanism that both legitimised the exploration of extreme or uncomfortable interventions and grounded inquiry in lived-like, relational contexts. On the one hand, it legitimised the exploration of otherwise implausible interventions that might be difficult to surface through conventional policy or design discussions, such as erasing shared interaction histories or memories to resolve conflict. While initially framed as exaggerated or speculative, these proposals surfaced substantive concerns, including how human control over shared memory could undermine agent independence and agency, an issue directly relevant to nowadays real-world human–AI accountability. At the same time, structured backstories involving named protagonists, shared routines, and relational stories ensured that contestation remained grounded in lived-like, situated concerns. This helped participants balance imaginative exploration with situated reasoning, preventing discussion from drifting into abstract or detached speculation. As P3 reflected, even when ideas felt \emph{“futuristic”} or \emph{“silly,”} the backstories enabled them to reason through consequences in a grounded and realistic way: \emph{“we argue about the pros and cons, but in a very realistic manner[…] this grounded type of logic helps us.”}

\section{Conclusion}

Rather than functioning as open-ended speculation, our method provides a shared methodological scaffold that expands the space of exploration while constraining discussion within plausible human–technology relationships. This directly responds to \citet{Yang2024HCIPolicy}’s call to treat futuring as a space for methodological exploration in HCI–policy collaboration, addressing their critique that many existing futuring approaches struggle to translate provocation into policy-relevant reasoning.
Our work presents a preliminary methodological exploration and we invite HCI researchers to further refine, extend, and adapt the method across different domains.

\begin{acks}
This research was partially funded by the UTS Transdisciplinary School and the Australian Research Council through the ARC Discovery Project DP220102019, Shared-Space Interactions Between People and Autonomous Vehicles. We sincerely thank all workshop participants for their valuable contributions.
\end{acks}





















\bibliographystyle{ACM-Reference-Format}
\bibliography{sample-base}

@String{Computing = "Computing" }

@String{Computer = "{IEEE} Computer" }

@inproceedings{Yu2025IASDR,
  author    = {Yu, Xinyan and Tran, Tram Thi Minh and Hoggenmüller, Marius and Tomitsch, Martin},
  title     = {From Screen to Street: Insights for Autonomous Mobility Design from Movies},
  booktitle = {Proceedings of the International Association of Societies of Design Research (IASDR)},
  year      = {2025},
  
}

@inproceedings{workshop2025,
author = {Yu, Xinyan and Berrio Perez, Julie Stephany and Hoggenm\"{u}ller, Marius and Tomitsch, Martin and Tran, Tram Thi Minh and Worrall, Stewart and Ju, Wendy},
title = {The UnScripted Trip: Fostering Policy Discussion on Future Human–Vehicle Collaboration in Autonomous Driving Through Design-Oriented Methods},
year = {2025},
isbn = {9798400720147},
publisher = {Association for Computing Machinery},
address = {New York, NY, USA},
url = {https://doi.org/10.1145/3744335.3749144},
doi = {10.1145/3744335.3749144},
booktitle = {Adjunct Proceedings of the 17th International Conference on Automotive User Interfaces and Interactive Vehicular Applications},
pages = {314–317},
numpages = {4},
location = {
},
series = {AutomotiveUI Adjunct '25}
}

@book{disalvo2015adversarialdesign,
  title={Adversarial design},
  author={DiSalvo, Carl},
  year={2015},
  publisher={Mit Press}
}

@inproceedings{Lupetti2023Steering,
author = {Lupetti, Maria Luce and Cavalcante Siebert, Luciano and Abbink, David},
title = {Steering Stories: Confronting Narratives of Driving Automation through Contestational Artifacts},
year = {2023},
isbn = {9781450394215},
publisher = {Association for Computing Machinery},
address = {New York, NY, USA},
url = {https://doi.org/10.1145/3544548.3581194},
doi = {10.1145/3544548.3581194},
booktitle = {Proceedings of the 2023 CHI Conference on Human Factors in Computing Systems},
articleno = {601},
numpages = {20},
location = {Hamburg, Germany},
series = {CHI '23}
}

@article{Lazar2010InteractingPolicy,
author = {Lazar, Jonathan},
title = {INTERACTING WITH PUBLIC POLICYInteracting with public policy},
year = {2010},
issue_date = {January + February 2010},
publisher = {Association for Computing Machinery},
address = {New York, NY, USA},
volume = {17},
number = {1},
issn = {1072-5520},
url = {https://doi.org/10.1145/1649475.1649485},
doi = {10.1145/1649475.1649485},
journal = {Interactions},
month = jan,
pages = {40–43},
numpages = {4}
}

@online{Kessler2012Uber,
  author    = {Sarah Kessler},
  title     = {Uber: When Innovation Outpaces the Law},
  year      = {2012},
  url       = {https://www.fastcompany.com/3001169/uber-when-innovation-outpaces-law},
  note      = {Accessed 11 Dec 2025}
}

@inproceedings{Jackson2014PolicyKnot,
author = {Jackson, Steven J. and Gillespie, Tarleton and Payette, Sandy},
title = {The policy knot: re-integrating policy, practice and design in cscw studies of social computing},
year = {2014},
isbn = {9781450325400},
publisher = {Association for Computing Machinery},
address = {New York, NY, USA},
url = {https://doi.org/10.1145/2531602.2531674},
doi = {10.1145/2531602.2531674},
booktitle = {Proceedings of the 17th ACM Conference on Computer Supported Cooperative Work \& Social Computing},
pages = {588–602},
numpages = {15},
location = {Baltimore, Maryland, USA},
series = {CSCW '14}
}

@inproceedings{Gairola2025CenteringHarm,
author = {Gairola, Ritika},
title = {Centering Harm in Socio-technical Systems: Connecting Design, User Experience, and Policy},
year = {2025},
isbn = {9798400714801},
publisher = {Association for Computing Machinery},
address = {New York, NY, USA},
url = {https://doi.org/10.1145/3715070.3747333},
doi = {10.1145/3715070.3747333},
booktitle = {Companion Publication of the 2025 Conference on Computer-Supported Cooperative Work and Social Computing},
pages = {11–14},
numpages = {4},
location = {
},
series = {CSCW Companion '25}
}

@inproceedings{Lindley2015BackToFuture,
author = {Lindley, Joseph and Coulton, Paul},
title = {Back to the future: 10 years of design fiction},
year = {2015},
isbn = {9781450336437},
publisher = {Association for Computing Machinery},
address = {New York, NY, USA},
url = {https://doi.org/10.1145/2783446.2783592},
doi = {10.1145/2783446.2783592},
booktitle = {Proceedings of the 2015 British HCI Conference},
pages = {210–211},
numpages = {2},
location = {Lincoln, Lincolnshire, United Kingdom},
series = {British HCI '15}
}

@inproceedings{grand2010designfiction,
  title     = {Design Fiction: A Method Toolbox for Design Research in a Complex World},
  author    = {Grand, Samuel and Wiedmer, Michael},
  booktitle = {Proceedings of the Design Research Society International Conference 2010},
  editor    = {Durling, David and Bousbaci, Rachid and Chen, Lin and Gauthier, Pierre and Poldma, Tiiu and Roworth-Stokes, Sally and Stolterman, Erik},
  year      = {2010},
  pages     = {},
  address   = {Montreal, Canada},
  publisher = {Design Research Society},
  url       = {https://dl.designresearchsociety.org/drs-conference-papers/drs2010/researchpapers/47}
}

@article{Auger2013SpeculativeDesign,
author = {James Auger},
title = {Speculative design: crafting the speculation},
journal = {Digital Creativity},
volume = {24},
number = {1},
pages = {11--35},
year = {2013},
publisher = {Routledge},
doi = {10.1080/14626268.2013.767276},
URL = {   https://doi.org/10.1080/14626268.2013.767276

}

}

@book{dunne2013speculative,
  title     = {Speculative Everything: Design, Fiction, and Social Dreaming},
  author    = {Dunne, Anthony and Raby, Fiona},
  year      = {2013},
  publisher = {The MIT Press},
  address   = {Cambridge, MA},
  isbn      = {9780262019842}
}

@inproceedings{Tost2024futuringMachines,
author = {Tost, Jordi and Gohsen, Marcel and Schulte, Britta and Thomet, Fidel and Kuhn, Mattis and Kiesel, Johannes and Stein, Benno and Hornecker, Eva},
title = {Futuring Machines: An Interactive Framework for Participative Futuring Through Human-AI Collaborative Speculative Fiction Writing},
year = {2024},
isbn = {9798400705113},
publisher = {Association for Computing Machinery},
address = {New York, NY, USA},
url = {https://doi.org/10.1145/3640794.3665904},
doi = {10.1145/3640794.3665904},
booktitle = {Proceedings of the 6th ACM Conference on Conversational User Interfaces},
articleno = {42},
numpages = {7},
location = {Luxembourg, Luxembourg},
series = {CUI '24}
}

@inproceedings{Blythe2017ResearchFiction,
author = {Blythe, Mark},
title = {Research Fiction: Storytelling, Plot and Design},
year = {2017},
isbn = {9781450346559},
publisher = {Association for Computing Machinery},
address = {New York, NY, USA},
url = {https://doi.org/10.1145/3025453.3026023},
doi = {10.1145/3025453.3026023},
booktitle = {Proceedings of the 2017 CHI Conference on Human Factors in Computing Systems},
pages = {5400–5411},
numpages = {12},
location = {Denver, Colorado, USA},
series = {CHI '17}
}

@inproceedings{Martelaro2020WhatCouldGoWrong,
author = {Martelaro, Nikolas and Ju, Wendy},
title = {What Could Go Wrong? Exploring the Downsides of Autonomous Vehicles},
year = {2020},
isbn = {9781450380669},
publisher = {Association for Computing Machinery},
address = {New York, NY, USA},
url = {https://doi.org/10.1145/3409251.3411734},
doi = {10.1145/3409251.3411734},
booktitle = {12th International Conference on Automotive User Interfaces and Interactive Vehicular Applications},
pages = {99–101},
numpages = {3},
location = {Virtual Event, DC, USA},
series = {AutomotiveUI '20}
}

@article{Bagloee2016AVPolicy,
  title   = {Autonomous vehicles: challenges, opportunities, and future implications for transportation policies},
  author  = {Bagloee, Saeed Asadi and Tavana, Madjid and Asadi, Mohsen and Oliver, Tracey},
  journal = {Journal of Modern Transportation},
  year    = {2016},
  volume  = {24},
  number  = {4},
  pages   = {284--303},
  issn    = {2196-0577},
  doi     = {10.1007/s40534-016-0117-3},
  url     = {https://doi.org/10.1007/s40534-016-0117-3}
}

@inproceedings{Kuo2025policycraft,
author = {Kuo, Tzu-Sheng and Chen, Quan Ze and Zhang, Amy X. and Hsieh, Jane and Zhu, Haiyi and Holstein, Kenneth},
title = {PolicyCraft: Supporting Collaborative and Participatory Policy Design through Case-Grounded Deliberation},
year = {2025},
isbn = {9798400713941},
publisher = {Association for Computing Machinery},
address = {New York, NY, USA},
url = {https://doi.org/10.1145/3706598.3713865},
doi = {10.1145/3706598.3713865},
booktitle = {Proceedings of the 2025 CHI Conference on Human Factors in Computing Systems},
articleno = {805},
numpages = {24},
location = {
},
series = {CHI '25}
}

@article{FAGNANT2015Policy,
title = {Preparing a nation for autonomous vehicles: opportunities, barriers and policy recommendations},
journal = {Transportation Research Part A: Policy and Practice},
volume = {77},
pages = {167-181},
year = {2015},
issn = {0965-8564},
doi = {https://doi.org/10.1016/j.tra.2015.04.003},
url = {https://www.sciencedirect.com/science/article/pii/S0965856415000804},
author = {Daniel J. Fagnant and Kara Kockelman}
}

@inproceedings{Yang2024HCIPolicy,
author = {Yang, Qian and Wong, Richmond Y. and Jackson, Steven and Junginger, Sabine and Hagan, Margaret D. and Gilbert, Thomas and Zimmerman, John},
title = {The Future of HCI-Policy Collaboration},
year = {2024},
isbn = {9798400703300},
publisher = {Association for Computing Machinery},
address = {New York, NY, USA},
url = {https://doi.org/10.1145/3613904.3642771},
doi = {10.1145/3613904.3642771},
booktitle = {Proceedings of the 2024 CHI Conference on Human Factors in Computing Systems},
articleno = {820},
numpages = {15},
location = {Honolulu, HI, USA},
series = {CHI '24}
}

@inproceedings{Spaa2019Boundaries,
author = {Spaa, Anne and Durrant, Abigail and Elsden, Chris and Vines, John},
title = {Understanding the Boundaries between Policymaking and HCI},
year = {2019},
isbn = {9781450359702},
publisher = {Association for Computing Machinery},
address = {New York, NY, USA},
url = {https://doi.org/10.1145/3290605.3300314},
doi = {10.1145/3290605.3300314},
booktitle = {Proceedings of the 2019 CHI Conference on Human Factors in Computing Systems},
pages = {1–15},
numpages = {15},
location = {Glasgow, Scotland Uk},
series = {CHI '19}
}

@inproceedings{Yang2023Towards,
author = {Yang, Qian and Wong, Richmond Y. and Gilbert, Thomas and Hagan, Margaret D. and Jackson, Steven and Junginger, Sabine and Zimmerman, John},
title = {Designing Technology and Policy Simultaneously: Towards A Research Agenda and New Practice},
year = {2023},
isbn = {9781450394222},
publisher = {Association for Computing Machinery},
address = {New York, NY, USA},
url = {https://doi.org/10.1145/3544549.3573827},
doi = {10.1145/3544549.3573827},
booktitle = {Extended Abstracts of the 2023 CHI Conference on Human Factors in Computing Systems},
articleno = {343},
numpages = {6},
location = {Hamburg, Germany},
series = {CHI EA '23}
}

@article{Sandhaus2023PrototypingAVCityPolicy,
  title   = {Towards Prototyping Driverless Vehicle Behaviors, City Design, and Policies Simultaneously},
  author  = {Sandhaus, Hauke and Ju, Wendy and Yang, Qian},
  journal = {arXiv preprint arXiv:2304.06639},
  year    = {2023},
  url     = {https://doi.org/10.48550/arXiv.2304.06639},
  doi     = {10.48550/arXiv.2304.06639}
}

@article{Mauri2024PolicySandboxing,
author = {Mauri, Andrea and Hsu, Yen-Chia and Verma, Himanshu and Tocchetti, Andrea and Brambilla, Marco and Bozzon, Alessandro},
title = {Policy Sandboxing: Empathy As An Enabler Towards Inclusive Policy-Making},
year = {2024},
issue_date = {November 2024},
publisher = {Association for Computing Machinery},
address = {New York, NY, USA},
volume = {8},
number = {CSCW2},
url = {https://doi.org/10.1145/3686908},
doi = {10.1145/3686908},
journal = {Proc. ACM Hum.-Comput. Interact.},
month = nov,
articleno = {369},
numpages = {42}
}

@article{braun2019reflecting,
  title={Reflecting on reflexive thematic analysis},
  author={Braun, Virginia and Clarke, Victoria},
  journal={Qualitative research in sport, exercise and health},
  volume={11},
  number={4},
  pages={589--597},
  year={2019},
  publisher={Taylor \& Francis}
}


\end{document}